
\magnification=1100 \def\wc{\hangindent=4em \hangafter=1 \noindent}
\baselineskip 12pt \parskip 3pt \null 
\headline={\ifnum\pageno=1 \hfil\hbox{
} \else\hfil\tenrm--\ \folio\ --\hfil\fi}
\footline={\hfil}

\def\la{\mathrel{\hbox{\rlap{\hbox{\lower4pt\hbox{$\sim$}}}\hbox{$<$}}}}
\def\ga{\mathrel{\hbox{\rlap{\hbox{\lower4pt\hbox{$\sim$}}}\hbox{$>$}}}}
\def\gram{\hbox{g}}
\def\erg{\hbox{erg}}
\def\keV{\hbox{keV}}
\def\MeV{\hbox{MeV}}
\def\sec{\hbox{s}}
\def\day{\hbox{day}}

\def\Hz{\hbox{Hz}}
\def\MHz{\hbox{MHz}}
\def\GHz{\hbox{GHz}}

\def\mJy{\hbox{mJy}}
\def\cm{\hbox{cm}}
\def\km{\hbox{km}}

\def\Gpc{\hbox{Gpc}}

\def\tco{t_{co}}
\def\tE{t_{\oplus}}
\def\tEm{t_{\oplus, m}}
\def\nucm{\nu_{co, m}}
\def\nuE{\nu_\oplus}

\def\FEm{F_{\nu, \oplus, m}}
\def\ee{{\cal E}}
\def\eemin{{\cal E}_{\rm min}}

\def\eemax{{\cal E}_{\rm max}}


\centerline{\bf RADIO TRANSIENTS FROM GAMMA-RAY BURSTERS}
\vskip 0.5cm
\centerline{Bohdan Paczy\'nski and James E. Rhoads}
\vskip 0.5cm
\centerline{Princeton University Observatory, Princeton, NJ 08544-1001}
\vskip 0.5cm
\centerline{I: bp@astro.princeton.edu~~and~~I: rhoads@astro.princeton.edu}
\vskip 0.5cm
\centerline{\it Received 1993 April 30; .................................}
\vskip 1.0cm
\centerline{ABSTRACT}
\vskip 0.5cm

The rapid time variability of gamma-ray bursts
implies the sources are very compact, and the peak luminosities are
so high that some matter must be ejected at ultra-relativistic speeds.
The very large Lorentz factors of the bulk flow are also indicated by
the very broad and hard spectra.  It is natural to expect that when the
relativistic
ejecta interact with the interstellar matter a strong synchrotron radio
emission is generated, as is the case with supernova remnants and
radio galaxies.  We estimate that the strongest gamma-ray bursts may be
followed by radio transients with peak fluxes as high as 0.1 Jy.
The time of peak radio emission depends on the distance scale; it is
less than a minute if the bursts are in the galactic halo, and about
a week if the bursts are at cosmological distances.

\vskip 0.3cm
\centerline{ {\it Subject headings:} cosmology: theory --- gamma rays:
bursts --- radio continuum: general --- relativity }

\vfill\eject
\centerline{1. INTRODUCTION}
\vskip 0.5cm

The isotropic distribution of gamma-ray bursts over the sky and
non-uniform distribution in distance (Meegan et al. 1992) indicates
that we are located at or near the center of a spherical and bound
distribution of sources.  This is easiest to understand if the sources
are at cosmological distances (Paczy\'nski 1991, Dermer 1992, Mao \&
Paczy\'nski 1992, Piran 1992).  The inferred peak luminosity is
$ \sim 10^{51} ~ \hbox{erg} ~ \sec^{-1} > 10^{17} ~ L_{\odot} $,
which is in excess of Eddington luminosity for any object with
$ M < 10^{13} ~ M_{\odot} $, and is highly super-Eddington for any
object that might generate a variability on a sub-millisecond time
scale.  This implies that the radiation pressure must drive an
ultra-relativistic wind (Paczy\'nski 1990, and references therein).
This conclusion remains valid even if the radiation is beamed into a
small solid angle.  The observed spectra are very broad and
non-thermal, extending to at least $ 10 ~ \MeV $ and in some cases
beyond $ 100 ~ \MeV $ and showing no evidence for a pair creation
cut-off at $ \sim 511 ~ \keV $ (Schaefer et al. 1992, and references
therein).  This is independent evidence for ultra-relativistic outflow
(Goodman 1986, Paczy\'nski 1986, Fenimore et al. 1992, 1993).

There are two other types of gamma-ray sources with spectra as broad
as burst spectra: some radio pulsars (Ruderman \& Cheng 1988)
and some AGNs (Hartman et al 1992, Dermer \& Schlickeiser 1992.)
Both classes of object are also
strong radio-emitters.  There may be a very general
reason for this, as both radio and gamma-ray emission require the
presence of relativistic particles.  Therefore, it is natural to expect
that gamma-ray bursts may give rise to radio emission as well (Paczy\'nski
1992).

There is plenty of evidence that all explosive events in astrophysics
give rise to synchrotron radio-emission when the ejecta interact with
the ambient medium.  Radio jets generated by active galactic nuclei
create `hot spots' when they are stopped by extragalactic matter,
and the products of the collision create giant radio-lobes, which outshine
the core sources at low frequences.  Supernova remnants are also
powerful radio-sources.

The most spectacular radio supernova known to date is SN 1986J in NGC 891,
at a distance of $ \rm \sim 12 ~ Mpc $.  The peak intensity
was $ \rm \sim 100 ~ mJy $ over the frequency range 1--10 GHz
(Rupen et al 1987) and the emission lasted for a few years.
The total radio energy generated by the supernova was
$ \rm E_{radio} \approx 10^{46} ~ erg $, and a surface brightness of
$ \rm T_b \approx 5 \times 10^{11} ~ K $ was reached, close to the Compton
limit of $ \rm 10^{12} ~ K $ (Kellerman \& Pauliny-Toth 1969).
The kinetic energy of supernova ejecta is typically
$ \rm E_{kin} \approx 10^{51} ~ erg $, hence the radio efficiency of
SN 1986J was $ \rm \sim 10^{-5} $.
We note that a ten minute observation
with the Very Large Array (VLA) reaches sensitivity limits $ \rm \sim $
(0.08, 0.1, 0.06, 0.2, 0.3) mJy at frequencies (1.4, 5.0, 8.4, 15, 23) GHz,
respectively (Very Large Array Observational Status Summary, 26 May 1993,
page 9, NRAO Publication).  Therefore, a supernova like SN 1986J is
detectable out to a distance $ \rm d \approx 0.4 ~ Gpc $.

A `typical' gamma-ray burst radiates $ \sim 10^{52} $ ergs in the 20
keV -- 2,000 keV range, assuming spherical symmetry.  Adopting $ \sim
10\% $ gamma ray efficiency, the total energy in the burst may be as
large as $ E_0 \approx 10^{53} $ ergs.  If the ejecta of some bursts are
as efficient radio emitters as SN 1986J then they may be detectable with
the VLA out to 1 Gpc and beyond.

In the following section we make the case for radio emission from gamma-ray
bursters somewhat more quantitative, and in the last section we discuss
the observable consequences of our scenario.

\vskip 0.5cm
\centerline{2. RADIO FIREBALL}
\vskip 0.5cm

There are $ \sim 40 $
gamma ray bursts per year per $ 4 \pi $ steradians out to a redshift
$ z \approx 0.2 $ (Mao \& Paczy\'nski 1992).  Every year the positions
for a handful of the strongest bursts are determined with
$ \sim 1 $ arc minute accuracy using the interplanetary network (IPN)
(Cline et al. 1992, and references therein).  A typical
distance to these bursts is likely to be $ 0.5 ~ \Gpc $ or so.

The following sequence of physical conditions seems to be fairly
general, and does not depend on the mechanism responsible for
the primary energy source or on the nature of the bursting object.  The
energy density at the source is so enormous that the optical depth to
all elementary processes is very high and the conditions are close to
local thermodynamic equilibrium, which implies that the gamma-ray
spectrum in the region is close to blackbody form (Goodman 1986,
Paczy\'nski 1986).  However, the observed spectra are very broad and
very different from any blackbody (Schaefer et al. 1992), and therefore
the observed emission cannot be due to the original fireball.
Fortunately, even a small `baryon loading' is sufficient to convert
almost all the original energy into kinetic energy of the relativistic
wind (Paczy\'nski 1990, Shemi \& Piran 1990).  In the observer's frame
the fireball looks like a thin spherical shell expanding
ultra-relativistically (Blandford \& McKee 1976, Shemi \& Piran 1990).
The shell cools rapidly because of adiabatic expansion.

The absence of the pair creation cut-off in the observed spectra implies
that the bulk Lorentz factor at the time of gamma-ray emission
is very large, $ \Gamma_{\gamma} \ge 10^2 $ (Fenimore et al 1992, 1993).
The initial bulk Lorentz factor of the fireball $ \Gamma_0 $ can only
by larger.  The collision between the ejecta and circum-source matter,
and later interstellar and/or intergalactic matter, will gradually
lower the bulk Lorentz factor, while converting the bulk
kinetic energy into random energy of relativistic particles as
described by Rees and M\'esz\'aros (1992) and M\'esz\'aros and Rees (1993)
in their model of gamma-ray bursts.  This `randomized' energy is partly
radiated away and partly used up for further expansion, i.e. it is
re-converted into kinetic energy of the bulk flow.  Ultimately, after a
long enough time interval, all the initial energy is either radiated away
or `wasted' in the Hubble expansion of the universe.

Assuming that the emission is incoherent, one can estimate the
synchrotron and inverse Compton radiation from the fireball (van der
Laan 1966, Rees 1967, Gould 1979).  There is an upper limit $ T_B \la
10^{12} ~ K $ to the brightness temperature of an incoherent
synchrotron source (Kellerman \& Pauliny-Toth 1969, Kellerman 1974).
In the observer's frame this limit is increased by the Lorentz factor
of the bulk flow.  Beyond this limit the energy density in radiation
($u_{rad}$) exceeds that in magnetic fields ($u_{_B}$), resulting in the
so-called ``Compton catastrophe,'' where the emitting electrons lose
energy very rapidly through inverse Compton scattering.  Such
conditions are likely to occur early in the evolution of the blast
wave, while all energy densities are still very high.  If they do
occur, the resulting inverse Compton emission may produce the observed
gamma-ray bursts - a scenario similar to the model of Rees and M\'esz\'aros.
When the expansion continues and the energy density falls below some critical
level, synchrotron emission dominates over inverse Compton losses.
This is the regime we are interested in.

The following analysis will be done assuming spherical symmetry of the
fireball.  We begin with the injection of energy $ E_0 $ into a small
volume ($ r_0 \la 300 \km $) with the initial rest mass $ M_0 $.  The
initial fireball quickly evolves into a thin shell expanding with bulk
Lorentz factor $ \Gamma_0 = E_0/M_0c^2 $.  By the time it has expanded
to $\sim \Gamma_0 r_0$, it becomes cold in the comoving
frame, i.e. the internal Lorentz factor $\gamma_i \approx 1$.
We assume that the ambient medium is at rest and has a uniform mass
density $\rho$.  As the blast wave expands, this material is swept up
and mixes with the shell, which is reheated and decelerated in the
process.
Parametrizing the swept up rest mass by $M = M_0(1+f)$, we can write
energy and momentum conservation as
$$
 \Gamma_0 + f = (1 + f) \gamma_i \Gamma , \eqno(1)
$$
$$
 \Gamma_0 \beta_0 = (1 + f) \gamma_i \Gamma \beta , \eqno(2)
$$
where $\Gamma = (1 - \beta^2)^{-1/2}$ is the bulk Lorentz factor,
$\beta = v/c$ as usual, and $ \gamma_i $ is the internal Lorentz factor
of the particles moving randomly within the expanding shell.  Solving these
gives $\Gamma$ and $\gamma_i$ as functions of $f$:
$$
 \beta = { \Gamma_0 \beta_0 \over \Gamma_0 +f } ,   ~~~~~~~
 \Gamma = ( 1 - \beta^2)^{-1/2} =
{ \Gamma_0 +f \over \sqrt{ 1 + 2 \Gamma_0 f + f^2 } } ~ ,  \eqno(3)
$$
$$
\rm \gamma_i = { \Gamma_0 + f \over (1+f) \Gamma } =
{ \sqrt{ 1 + 2 \Gamma_0 f + f^2 } \over 1+f } ~ , ~~~~~~~
\beta_i = (1 - \gamma_i^{-2})^{1/2} ~~.  \eqno(4)
$$

There are three different measures of time which are relevant.
The first, $t$, is measured in the rest frame of the burster, so that
$$
r \approx ct, ~~~~~~~~~
M = M_0(1+f) = M_0 ~ + ~ { 4 \pi \over 3 } (ct)^3 \rho , \eqno(5)
$$
where $ r $ is the radius of the shell.  The second, $\tco$, is measured
in the frame comoving with the shell, so that
$$
\tco = \int_0^t { d\,t \over \Gamma} .  \eqno(6)
$$
The third, $\tE$, is measured in the terrestrial observer's frame and must
account for apparent superluminal motion, so that
$$
\tE = t - { r \over c } \cos \theta .  \eqno(7)
$$
We may write these in terms of $f$ as
$$
t = {1 \over c } \left( 3 M_0 \over 4 \pi \rho \right)^{1/3} f^{1/3} ,
\eqno(8)
$$
$$
\tco \approx {t \over \Gamma_0 } \left(1 + 0.32 \Gamma_0 f \right) ^{1/2}
{}~~~~ , ~~~~ 0 \le f \la \Gamma_0
\eqno(9)
$$
$$
\tE \approx {t \over 2 \Gamma_0^2} \left( 1 + 0.5 \Gamma_0 f \right)
{}~~~~ . ~~~~ 0 \le f \la \Gamma_0  \eqno(10)
$$
Here the approximate form for $\tco$ interpolates between results
accurate for $ f \ll \Gamma_0^{-1} $ and for $ \Gamma_0^{-1} \ll f \ll
\Gamma_0 $, while $\tE$ is calculated for $\theta = 0$, i.e. along the
line of sight.  It is convenient to develop our model first in terms of
$f$ and later transform to $\tE$.

The expanding shell as seen in its comoving frame will expand at a rate
comparable to the sound speed $c / \sqrt{3}$, so that the thickness is
$\Delta r \approx c \tco$.  The shell's surface is perpendicular to the
direction of expansion, so it is the same in the comoving and
stationary frames, $A = 4 \pi (c t )^2$.  Thus, in the frame co-moving
with the shell its volume is given as
$$
V \approx 4 \pi (ct)^2 (c \tco) \approx {3 M_0 \over \rho \Gamma_0} f
\left( 1 + 0.32 \Gamma_0 f \right) ^{1/2}  ~~ ,           \eqno(11)
$$
and its total energy as
$$
E_i = u_i V =
M_0 (1 + f) \gamma_i c^2 = M_0 c^2 \left( 1 + 2\Gamma_0 f + f^2 \right) ^{1/2}
{}~ , \eqno(12)
$$
where $ u_i $ is the total energy density.

The expanding shell is reheated to relativistic energies and the
internal Lorentz factor may be as large as
$ \gamma_{i,max} \approx ( \Gamma_0 / 2)^{1/2} $ (cf. eq. 4).
These are ideal conditions for synchrotron emission (cf. Rees \&
M\'esz\'aros 1992, M\'esz\'aros \& Rees 1993, and references therein).
We take the electrons to have the energy distribution $N(\ee)
\propto \ee^2$ for $\eemin < \ee < \eemax$, i.e., a power law
distribution with equal energy per decade.  This is similar to electron
energy spectra inferred from synchrotron emission observed in a variety
of sources, and the choice of exponent 2 reduces the sensitivity of the
results to $\eemin$ and $\eemax$.  Working for the moment in the frame
of the shell we adopt the standard results for synchrotron emission from a
power law distribution of electrons (cf. Pacholczyk 1970, \S 3.4).
The spectrum takes the form $I_\nu
\propto \nu^{5/2} (1 - e^{-\tau})$ where $\tau$ is the optical
depth to synchrotron self-absorption.  For a power law exponent of 2 and
frequencies in the range $ \nu_c(\eemin) \la \nu \la \nu_c(\eemax) $,
we have in the frame comoving with the shell
$$
\tau = 0.35 ~ w^{-1} \left( {\Delta r \over c }
\right) u_e u_{_B} \left( { \nu \over 5 \GHz } \right)^{-3} ~~\hbox{cgs}
\eqno(13)
$$
where $ w \equiv \log_{10}({\eemax / \eemin})$, $\Delta r$ is the thickness
of the emitting slab, $ u_e $ and $ u_{_B} $ are the electron and the magnetic
energy densities, and $\nu_c$ is the characteristic frequency for synchrotron
emission of a single electron, given by $\nu_c(\gamma_e) \approx 5 \times
10^5 \left(\ee / [m_e c^2] \right)^2 (B / \hbox{gauss})$.  It follows that
the spectrum peaks at the
frequency $\nu_{m,co} $ for which $\tau = 0.35$, rising as $\nu^{5/2}$
at lower frequencies and falling off as $\nu^{-0.5}$ at higher ones.
The peak specific intensity is
$$
I_{\nu, m,co} = 4.3 \times 10^{-7} \left(B \over \hbox{gauss}\right)^{-1/2}
\left(\nu_{m,co} \over 5 \GHz \right)^{5/2} ~~~~
\left[ { \erg \over \cm^2 \sec \,\Hz\, \hbox{Steradian} } \right] . \eqno(14)
$$

We do not know the magnetic energy density $ u_{_B} = B^2/8 \pi $
or the electron energy density $ u_e $, so we express them in
terms of dimensionless parameters,
$$
u_{_B} = \xi_{_B} u_i , ~~~~~~ u_e = \xi_e u_i , \eqno(15)
$$
and conservatively guess that $ \xi_e \sim 0.01 $,
well below the equipartition value - the maximum plausible range
is $ (m_e / m_p) \la \xi_e \la 0.5$; given enough time the magnetic
field tends to approach equipartition but it is not likely to be
that strong from the beginning of the sweeping up phase, so it seems
reasonable to take $ \xi_{_B} \sim 0.01 $ as a representative value.
We can constrain the range over which the electron distribution obeys
$N(\ee) \propto \ee^2$ by assuming that inverse Compton losses enforce
$u_{rad} \la u_{_B}$. This gives an upper bound on $w$ typically in the
range $2$ to $8$, with lower $\xi_e$ allowing higher $w$.

The following is a rather tedious task.  First we have to express
$ \tau  $, $ \nucm $ and $ I_{\nu , m,co} $ from eqs. (13)
and (14) in terms of the fireball parameters given in eqs. (3-5).  Next,
we have to transform to the frame of the Earth-bound observer.
The relativistic transformations of the frequency and the specific
intensity are
$$
\nu_{\oplus} = (1 + \beta) \Gamma \nu_{co} \approx 2 \Gamma \nu_{co} ~,
{}~~~~~~
I_{\nu, \oplus} = (1 + \beta)^3 \Gamma^3 I_{\nu, co} \approx 8 \Gamma^3 I_{\nu,
co} ~ .
\eqno(16)
$$
To go from the specific intensity to a flux density we multiply by the
apparent size of the source,
$$
\FEm \approx { \pi (ct)^2 \over d^2 \Gamma^2 } 8 \Gamma^3 I_{\nu, co, m}
{}~ .  \eqno(17)
$$
The relation between $f$ and $\tE$ is given in eq. (10).
We will restrict our attention to the regime $ \Gamma_0^{-1} \la
f \la \Gamma_0 $, so that $ \tE \approx ( f t ) / ( 4 \Gamma_0 ) $.

After some algebra we find that the time from the beginning of the
fireball to the observed peak in the flux at frequency $\nuE$ is
$$
\tEm \approx 12 \day ~ C_t ~
\left( {E_0 \over 10^{53} \erg}\right) ^{1/2}
\left( {\rho \over 10^{-24} \gram ~ \cm^{-3}} \right)^{1/2}
\left( { \nuE \over 5 \GHz } \right)^{-3/2}
\eqno(18a)
$$
$$
\approx 21 \day ~ C_t ~ { \xi_{\gamma} \over 0.1 } ~
\left({d \over 0.5 \Gpc } \right)
\left( { S \over 10^{-3} \erg ~ \cm^{-2} }\right)^{1/2}
\left( {\rho \over 10^{-24} } \right)^{1/2}
\left( { \nuE \over 5 \GHz } \right)^{-3/2} ,
\eqno(18b)
$$
and that the corresponding peak flux is
$$
\FEm \approx 56 \mJy ~ C_F ~
\left({d \over 0.5 \Gpc } \right)^{-2}
\left({E_0 \over 10^{53} } \right)^{7/8}
\left(\rho \over 10^{-24}\right)^{1/8}
\left( { \nuE \over 5 \GHz } \right)^{5/8}
\eqno(19a)
$$
$$
\approx 150 \mJy ~ C_F ~ { \xi_{\gamma} \over 0.1 } ~
\left({d \over 0.5 \Gpc } \right)^{-1/4}
\left( { S \over 10^{-3} }\right)^{7/8}
\left(\rho \over 10^{-24}\right)^{1/8}
\left( { \nuE \over 5 \GHz } \right)^{5/8}
\eqno(19a)
$$
where
$$
C_t \equiv \left( { \xi_e \over 0.01 } ~ { \xi_{_B} \over 0.01 } ~
{ 5 \over w } \right)^{1/2}
{}~~~ , ~~~~~~~~~
C_{_F} \equiv \left( { \xi_e \over 0.01 } \right)^{5/8}
\left( { \xi_{_B} \over 0.01 } \right)^{9/24}
\left( { 5 \over w } \right)^{5/8} ~~ , \eqno(20)
$$
are dimensionless coefficients of order unity, $ S $ is the burst fluence
as measured at Earth ($ E_{\gamma} = 4 \pi d_0^2 S $), and
$ \xi_{\gamma} = E_{\gamma} / E_0 $ is the
fraction of fireball's energy radiated in gamma-rays in
in the range $ 20 - 2,000 ~ \keV $.
The bulk Lorentz factor at the time of peak emission at $\nuE$ is
$$
\Gamma_{radio} \approx 3 ~ C_t^{-3/8}
\left( E_0 \over 10^{53} \right)^{-1/16}
\left(\rho \over 10^{-24}\right)^{-5/16}
\left(\nuE \over 5 \GHz \right)^{9/16} ~~ .
\eqno(21)
$$
By combining equations~(18) and~(19) with the spectrum we find that at a
fixed frequency the flux rises as $\tE^{5/4}$ prior to $\tEm$ and
declines as $\tE^{-3/4}$ thereafter, so that the total time above half
maximum is about $2 \tEm$.  We can also use (18) and (19) to calculate the
peak brightness temperature $T_{b, m} = c^2 I_{\nu, co, m} / (2 \nucm^2 k) $.
The result is approximately $10^{11} ~ K$ and is extremely insensitive
to all parameters of the model.

\vskip 0.5cm
\centerline{3. DISCUSSION}
\vskip 0.5cm

Observational upper limits to the radio emission from GRBs are
available in the literature for timescales $\tE \la 10\ \hbox{hours}$
(Baird et al 1975; Cortiglioni et al 1981) and for $\tE \approx 5~
\hbox{years}$ (Schaefer et al 1989).  Baird et al (1975) used $\sim$
full sky coverage at frequencies $\nuE \le 151 \MHz$ to watch for
strong radio bursts ($F_\nu \ga 100\>\hbox{kJy}$, duration $\le 100~
\sec$) and later checked for events that had occurred within 10 hours of
recorded GRBs.  Cortiglioni et al (1981) used a similar method at $151$
and $408 \MHz$, attaining a somewhat higher sensitivity ($F_\nu \ga
1\>\hbox{kJy}$) and looking for closer time coincidence ($\le 10~
\hbox{minutes}$) with GRBs.  Schaefer et al (1989) used the VLA at
$15$, $5$, and $1.4 \GHz$ to conduct a deep search for quiescent
emission in IPN GRB error boxes, with sensitivity from $0.1$ to $0.8
\mJy$.  Other potentially relevant searches have been done by Vaughan
and Large (1987) and by Amy et al (1989).  However, none of these
studies found anything above a normal rate of background events.
Other efforts are underway, but there have been no confirmed detections
of radio emission from GRBs, nor are we aware of published limits on a
timescale of a few days.  Hanlon, Bennett, and Spoelstra reported a
candidate detection ($76 \mJy$ at $0.6 \GHz$, $< 1 \mJy$ at $5 \GHz$)
three days after GRB 930309 (IAU Circular 5749; cf.\ also IAU Circulars
5750, 5755, 5763, and 5764 [1993]) but IPN data showed the radio source
and GRB positions to be inconsistent.

None of these experiments provides serious restrictions on our scenario.
If gamma-ray bursts are at cosmological distances then we expect the
peak of the radio emission to follow the burst within a week or so, and for
the strongest bursts to be up to 0.1 Jy at frequencies of
a few GHz.  It is clear that a dedicated search
with the VLA or other sensitive radio instruments should follow accurate
position determinations from the Interplanetary Network
as soon as possible.  It would be best
to observe the rise as well as the fall of the radio transient as this
would provide very good diagnostics for the burst environment, and
in particular for the distance scale.  If the bursts
are in the galactic halo, their energies are likely to be $ \sim 10^{43} $
ergs rather than $ 10^{53} $ ergs, and so the time scale of the transient
would be $ \sim 10 $ seconds, i.e. practically simultaneous with the gamma-ray
bursts (cf. eq. 18).  The peak radio flux depends
weakly on the distance scale (cf. eq. 19), still the galactic halo radio
transients might be as strong as $ \sim $ 1 Jy.

It should be noted that there is one major uncertainty in our estimate:
the gamma-ray burst may be strongly beamed, perhaps to a solid angle as
small as $ \pi \Gamma_{\gamma}^2 ~ \sim 10^{-5} $.
In this case the energy required to power the burst may be reduced
to a value as small as $ E_0 \sim 10^{48} \erg $
without affecting the observable properties of the gamma-ray burst,
but strongly reducing the radio power which peaks at $ \Gamma_{radio}
\approx 3 $ (cf. eq. 21).  The modest amount of the radio beaming may
bring the equivalent fireball energy up to $ E_0 \sim 10^{49} ~ \erg $,
but our estimate of the peak radio power is reduced by a factor
$ \sim 10^4 $ making the radio transient undetectable (cf. eq. 19).
To maximize the likelihood of detecting a radio transient one should
observe at the highest frequency possible, as the total radio power
increases with frequency (cf. eq. 19) and the radio beaming factor increases
as $ \Gamma_{radio}^2 $ (cf. eq. 21).  Of course, the higher the radio
frequency the more rapid the radio transient (cf. eq. 18), and the more
rapidly must the VLA observation follow the gamma-ray burst.

\vskip 0.4cm

This project was supported by the NASA grant NAG5-1901 and the NSF grant
AST90-23775.  Part of this work was completed when BP was a Distinguished
VITA Visitor at the University of Virginia, and he would like to acknowledge
the support and hospitality of the Virginia Institute of Theoretical
Astronomy, and stimulating discussion with Dr. R. Chevalier.

\vfill\eject
\centerline{REFERENCES}
\vskip 0.5cm

\wc{Amy, S. W., Large, M. I. \& Vaughan, A. E. 1989,  Proc. Astron. Soc.
Australia, 8, 172 \hfill}

\wc{Baird, G. A., et al. 1975, Ap. J. Lett., 196, L11 \hfill}

\wc{Blandford, R. D., \& McKee, C.F. 1976, Physics of Fluids, 19, 1130 \hfill}

\wc{Cline, T. L. et al. 1992, in {\it The Compton Observatory Science
Workshop}, p. 60 (Eds.: C. R. Shrader, N. Gehrels, \& B. Dennis, NASA)
\hfill}

\wc{Cortiglioni, S., Mandolesi, N., Morigi, G., Ciapi, A., Inzani, P.,
\& Sironi, G. 1981, Astrophysics and Space Science 75, 153 \hfill}

\wc{Dermer, C. D. 1992, Phys. Rev. Lett., 68, 1799 \hfill}

\wc{Dermer, C. D., \& Schlickeiser, R. 1992, Science, 257, 1642 \hfill}

\wc{Fenimore, E. E., Epstein, R. I., \& Ho, C. 1992, in {\it Gamma-Ray
Bursts}, AIP Conf. Proc. 265, p. 158 (Eds.: W. S. Paciesas \& G. J.
Fishman) \hfill}

\wc{Fenimore, E. E., Epstein, R. I., \& Ho, C. 1993, A\&A Suppl, 97, 59
\hfill}

\wc{Goodman, J. 1986, ApJ, 308, L47 \hfill}

\wc{Gould, R. J. 1979, Astronomy and Astrophysics, 76, 306  \hfill}

\wc{Hartman, R. C. et al. 1992, ApJ, 385, L1 \hfill}

\wc{International Astronomical Union Circulars 5749, 5750, 5755, 5763,
\& 5764  A.D. 1993 \hfill}

\wc{Kellerman, K. I. 1974, in {\it Galactic and Extragalactic Radio
Astronomy}, p. 320 (Eds.: G. L. Verschuur \& K. I. Kellerman)  \hfill}

\wc{Kellerman, K. I. \& Pauliny-Toth, I. I. K. 1969, ApJ, 155, L71 \hfill}

\wc{Mao, S. \& Paczy\'nski, B.  1992, ApJ, 388, L45 \hfill}

\wc{Meegan, C. A. et al. 1992, Nature, 355, 143 \hfill}

\wc{M\'esz\'aros, P. \& Rees, M. J  1993, ApJ, 405, 278  \hfill}

\wc{Pacholczyk, A. G. 1970, Radio Astrophysics, (W. H. Freeman \& Co.,
San Francisco)  \hfill}

\wc{Paczy\'nski, B. 1986, ApJ, 308, L43 \hfill}

\wc{Paczy\'nski, B. 1990, ApJ, 363, 218 \hfill}

\wc{Paczy\'nski, B. 1991, Acta Astron., 41, 257 \hfill}

\wc{Paczy\'nski, B. 1992, Princeton Observatory Preprint 463,
solicited by Science, and rejected \hfill}

\wc{Piran, T. 1992, ApJ, 389, L45 \hfill}

\wc{Rees, M. J. \& M\'esz\'aros, P. 1992, MNRAS, 258, 41p  \hfill}

\wc{Rees, M. J. 1967, MNRAS, 135, 345  \hfill }

\wc{Ruderman, M. \& Cheng, K. S. 1988, ApJ, 335, 306 \hfill}

\wc{Rupen, M. P. et al. 1987, AJ, 94, 61 \hfill}

\wc{Schaefer, B. E. et al. 1989, ApJ, 340, 455 \hfill}

\wc{Schaefer, B. E. et al. 1992, ApJ, 393, L51 \hfill}

\wc{Shemi, A. \& Piran, T. 1990, ApJ, 365, L55 \hfill}

\wc{van der Laan, H. 1966, Nature, 211, 1131  \hfill}

\wc{Vaughan, A. E., \& Large, M. I. 1987, Astrophysical Letters \&
Communications, 25, 159 \hfill}

\wc{Very Large Array Observational Status Summary, 26 May 1993,
NRAO Publication \hfill}

\vfill \end \bye